# Pressure dependence of Ce valence in CeRhIn$_5$

**Z E Brubaker[1,2], R L Stillwell[2], P Chow[3], Y Xiao[3], C Kenney-Benson[3], R Ferry[3], Z Jenei[4], R J Zieve[1] and J R Jeffries[4]**

[1] Physics Department, University of California, Davis, California, USA
[2] Materials Science Division, Lawrence Livermore National Laboratory, Livermore, California 94550, USA
[3] HPCAT, Geophysical Laboratory, Carnegie Institute of Washington, Argonne National Laboratory, Argonne, Illinois 60439, USA
[4] Physics Division, Lawrence Livermore National Laboratory, Livermore, California 94550, USA

E-mail: zbrubaker@ucdavis.edu

**Abstract**
We have studied the Ce valence as a function of pressure in CeRhIn$_5$ at 300 K and at 22 K using x-ray absorption spectroscopy in partial fluorescent yield mode. At room temperature, we found no detectable change in Ce valence greater than 0.01 up to a pressure of 5.5 GPa. At 22 K, the valence remains robust against pressure below 6 GPa, in contrast to the predicted valence crossover at P=2.35 GPa. This work yields an upper limit for the change in Ce-valence and suggests that the critical valence fluctuation scenario, in its current form, is unlikely.

# 1. Introduction

Heavy-fermion materials can qualitatively be understood by considering the Doniach phase diagram. [1] The ground state is determined by the relative strength of the Kondo interaction to the RKKY interaction, which can be continuously tuned via parameters such as magnetic field, pressure, and chemical substitution. In the RKKY dominated regime, these materials exhibit magnetic order, while the Kondo regime favors Fermi Liquid behavior. [2,3] Between these two regions, a quantum critical point (QCP) is often observed, with a V-shaped non-Fermi liquid region emanating from the QCP. It is near the QCP, that superconducting domes often emerge. Due to the proximity to magnetic order, magnetic fluctuations have been suggested to account for this superconducting state. [4] In some materials, however, a second dome emerges farther from the magnetic state, revealing distinct behavior and necessitating an alternate description. [5]

The Critical Valence Fluctuation (CVF) scenario has been proposed to elucidate the behavior of these distinct superconducting domes. Within this scenario, there should exist a sharp valence transition near the critical pressure of these domes. The CVF scenario has been suggested to account for some materials in both the Ce-122 and Ce-115 families. [6-8] With increasing pressure, $CeCu_2Si_2$ and $CeCu_2Ge_2$ reveal the two superconducting domes described above, whereas the Ce-115s are less clear. [9-11] While each of the Ce-115s display only a single dome with pressure, $Ce(Rh_{1-x}Ir_x)In_5$ reveals two domes with increasing Rh content. [12-15] Additionally, the Ce-115 family demonstrates many properties consistent with the CVF scenario -- such as heavy quasiparticles forming within the antiferromagnetic (AFM) state, a peak in resistivity and an abrupt change in Fermi surface – which suggest that the Ce-115 family may fit within the framework of the CVF scenario. [6]

As a test of the CVF scenario, the Ce valence in $CeCu_2Si_2$ has previously been studied, and was found to display a smooth valence change from 3.04 to 3.17, in contrast to the expected sharp crossover implied by the CVF scheme and the valence discontinuity associated with the gamma-alpha transformation in Ce metal. [16, 17] The Ce-valence of $CeCoIn_5$ and $CeIrIn_5$ has also been studied under pressure, both revealing at most subtle changes in valence. [18] While the CVF scenario was developed to explain low-temperature behavior, the temperature dependence of the valence crossover is unclear and likely material dependent. In the cases of $CeCu_2Si_2$ and the other Ce-115s, previous valence measurements were performed at T=14 K – 16 K, a temperature above both $T_c$ and $T_K$ in any of these materials. [19-21] However, critical behavior should be observable at $T \gg T_c$, so valence fluctuations should remain present even at higher temperatures. [2]

To probe the CVF model in $CeRhIn_5$, we have studied the Ce-valence under pressure at both T=300 K and T=22 K using x-ray absorption spectroscopy (XAS). The compound $CeRhIn_5$ is predicted to have a sharp valence crossover near P=2.35 GPa, where $T_c$ reaches its maximum value of 2.22 K. [7,8] While the Kondo temperature for dilute Ce in $LaRhIn_5$ is 0.15 K, the Kondo scale for a fully dense lattice of Ce in $CeRhIn_5$ is estimated to be 10-28 K, a $T_K$ near that of its 115 cousins, but also a temperature that is extremely challenging for pressure-dependent, synchrotron-based valence measurements due to excess attenuation from vacuum chamber windows, radiation shield apertures, diamonds, and gasket materials required for work at these extreme conditions. [22-25] The coherence temperature of $CeRhIn_5$ is found to be $T^* \approx 18K$ and increases with pressure. [26] Both the Kondo and coherence temperatures are thus comparable to the temperature at which the cryogenic measurements reported herein were performed.

# 2. Experimental Methods

High-quality $CeRhIn_5$ crystals were grown using an indium flux method reported elsewhere in the literature and were etched with 25% HCl. [27] The crystal structure was verified using powder x-ray diffraction, which indicated single-phase $CeRhIn_5$. Pressure was generated using a diamond anvil cell (DAC) with a beryllium gasket. Small crystals (<10 micron) were loaded into the sample chamber of the DAC using mineral oil as the pressure-transmitting medium. The pressure was measured via standard ruby fluorescence spectroscopy. [28]

XAS at the Ce L-III edge was performed at sector 16-IDD (HPCAT) of the Advanced Photon Source using partial fluorescence yield (PFY) from the L$\alpha$ emission line. The incident energy was scanned with a Si 111 fixed exit double crystal monochromator, and the L$\alpha$ emission was recorded using a) a single Si 400 analyzer for ambient-temperature measurements at T=300 K and b) three Si 400 analyzers for measurements with the DAC inside of a cryostat. The analyzer crystals were aligned to focus the Ce Lα emission onto a Pilatus detector. Low temperature work was performed in a helium flow cryostat, using ~250 micron kapton in the incident window and 50 micron beryllium in the exit window. The lowest stable temperature obtainable was T=22 K. Each x-ray absorption near edge structure (XANES) spectrum at a given pressure is the result of three or more individual energy scans, which were summed to obtain a satisfactory signal-to-noise ratio.

## 3. Results and Discussion

L-III XAS is sensitive to the valence of the Ce ions, because the 4f-states ($f^1$ and $f^0$) each cause different screening of the 2p3/2 core-level excitations. Owing to the large unoccupied density of states arising from the unfilled 5d-shell of the Ce ions, the Ce L-III XAS spectra present a prominent and distinct "white line" peak for each valence configuration. These distinct peaks will occur at different energies separated by approximately 8-12 eV, a splitting greater than the intrinsic core-hole lifetime broadening of the 3d-2p (Lα emission) final state in the PFY emission. Therefore, a hypothetical valence change from an $f^1$ to an $f^0$ configuration should result in a spectrum that clearly shifts such that the white line is centered about 8-12 eV higher in incident energy. By comparing the intensities of the $f^1$ and $f^0$ peaks, we can determine the Ce valence.

Figures 1a and 1b show the absorption spectra at each measured pressure for T=300 K and T=22 K respectively, which have been normalized to an edge jump of unity. The locations of the expected peaks for the $f^1$ and $f^0$ states are indicated, as well as the pre-edge feature (*), which has been interpreted as the $f^2$ peak in several recent papers. [16, 18] The pre-edge feature likely corresponds to the 2p → 4f quadrupole transition, as discussed in several resonant inelastic x-ray scattering (RIXS) and XANES measurements performed on other lanthanide compounds. [29-32]

The predicted change in Ce valence should be observed by a peak developing in the $f^0$ region above the critical pressure. At T=300 K, the $f^1$ peak decreases in amplitude with pressure, but the $f^0$ region remains unaffected. Part of this drop is due to the ambient pressure measurement being performed outside of the pressure cell, but there remains a small change associated with increasing pressures. The decrease in size of the $f^1$ peak is not indicative of a valence change, but is associated with a reduction in the unoccupied 5d-electron density of states owing to pressure-induced 6s-5d charge transfer. [33]

At T=22 K, the $f^1$ peak also decreases in intensity with pressure, though it is less consistent than at ambient pressure, likely due to the additional scattering sources in the cryogenic setup. There also appears to be an enhancement in the pre-edge region in the P=2.4 GPa curve, just above $P_c$ in this compound. However, given the similar glitch in the P=4.4 GPa data around 5.74 keV, this is likely an artifact of the experimental setup. The $f^0$ peak does not reveal any enhancement around $P_c$, though at P=6 GPa there exists a minor increase. We cannot rule out the possibility that this is caused by the noisy background, but because the energy coincides with the location of the $f^0$ peak and similar increases have been measured on other Ce-115 compounds [18], we have performed the analysis under the assumption that this corresponds to an $f^0$ peak.

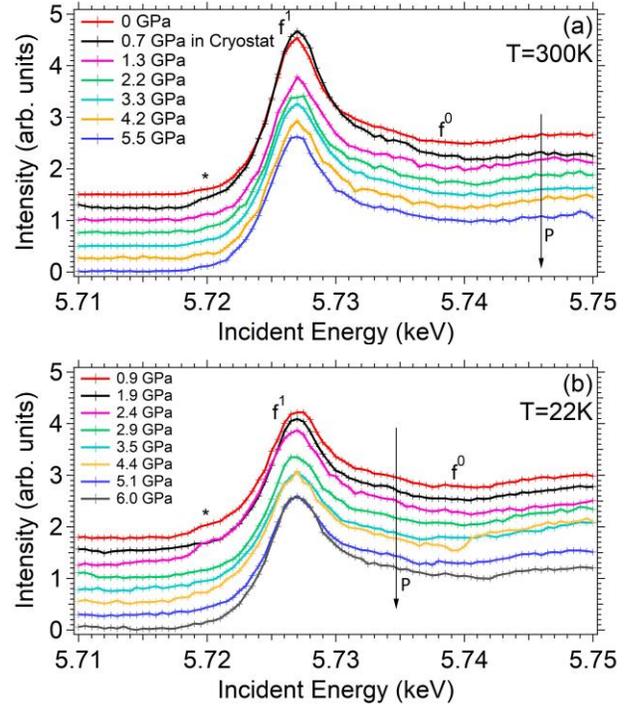

**Figure 1 (Color online) Absorption Spectra at (a) T=300 K and (b) T=22 K for CeRhIn$_5$ as a function of pressure.** Each spectrum is offset by 0.25. The pre-edge feature, *, develops significant fluctuations when measured with the cryostat in place, which are likely artifacts of the experimental setup. The P=6 GPa spectrum at T=22 K reveals a subtle increase at 5.738 keV, which has been interpreted as the $f^0$ peak.

Figure 2a shows our fit for the ambient pressure data at T=300 K. The $f^1$ and $f^0$ peaks are fit with a Gaussian and error function fixed at their respective absorption edges. In addition to these peaks, we added a Gaussian to account for the pre-edge feature, and an

$f^1$ shoulder centered at ~5.7314 keV. The $f^1$ shoulder has been shown to decrease along with the $f^1$ peak in CeCu$_2$Si$_2$ indicating that it is related to the same family of states as the $f^1$ peak, but this shoulder does not evolve with pressure in CeRhIn$_5$ and does not enter our analysis of the valence. [16] There is no detectable contribution of the $f^0$ peak at ambient pressure at T=300 K, yielding a valence of 3.00 (+.025 -0.00). Figure 2b shows the fit for the P=6 GPa measurements at T=22 K, which reveals a 3% contribution from the $f^0$ peak.

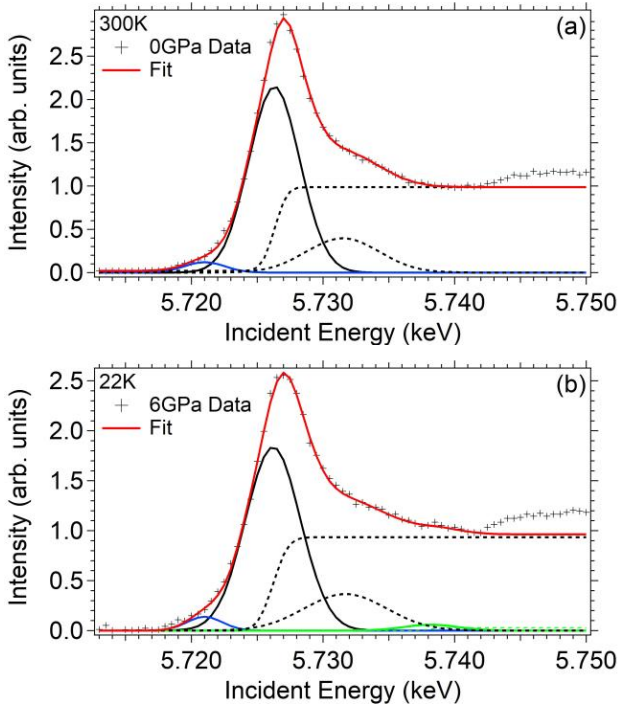

**Figure 2 (Color online) Example fit of (a) T=300 K, P=0 GPa data and (b) T=22 K, P=6 GPa data.** The solid lines correspond to absorption peaks that entered the analysis, while the dashed lines correspond to error functions associated with each valence peak and the gaussian accounting for the shoulder of the $f^1$ peak (~5.7314 keV), which did not enter the analysis. The blue curve corresponds to the pre-edge feature, centered at ~5.721 keV, the black curves correspond to those associated with the $f^1$ peak centered at ~5.726 keV, and the green curves correspond to those associated with the $f^0$ peak, centered at ~5.738 keV. Spectra were fit up to E=5.742 keV.

Similar analysis has been performed on each pressure curve, and the results are summarized in Figure 3, where the valence determination is overlaid on the P-T phase diagram. At T=300 K, the valence remains a constant n=3.00 up to the highest measured pressure of P=5.5 GPa. At low temperature, the valence remains unchanged up to P=5.1 GPa, in contrast to the expected valence crossover at P$_c$=2.35 GPa. Only at P=6 GPa is there a slight increase in valence, rising to 3.03 (+0.055 -.03). This subtle increase in the determined valence is unlikely to be related to the CVF scenario, not significantly distinct within error from a flat trend with pressure, and similar in magnitude to previous observations on other Ce-115 compounds. [18] The absence of any evidence of a valence crossover in previous work measuring CeCu$_2$Si$_2$, CeIrIn$_5$ and CeCoIn$_5$ [16,18], in addition to the current work measuring CeRhIn$_5$, suggests that the CVF scenario, in its current form, is unlikely.

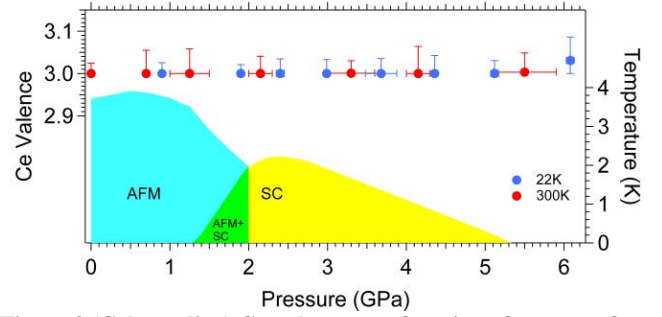

**Figure 3 (Color online) Ce-valence as a function of pressure for CeRhIn$_5$.** Phase Diagram modeled after [34]. The valence error was determined from counting statistics and standard error propagation using the calculated error from weighted fitting in Igor. Additionally, our analysis focuses on the $f^1$ and $f^0$ peaks, so it is not possible to have a valence of n<3, which accounts for the asymmetric error bars.

## 4. Conclusion

In summary, we have directly measured the valence at the Ce L-III edge in CeRhIn$_5$ under pressure using XAS-PFY at T=300 K and T=22 K. At T=300 K the valence remains constant at all measured pressures, while at T=22 K a slight increase in valence is observed at P=6 GPa. Neither of these measurements support the CVF scenario, and suggest that within our resolution no change in valence exists at the critical pressure. The magnetic fluctuation scenario is likely sufficient to explain superconductivity in the Ce-115s, though further work is required to understand the nature of superconducting domes far from magnetic order in the Ce-122s.

## Acknowledgements

The authors thank Per Soderlind for fruitful discussion. This work was performed under LDRD (Tracking Code 14-ERD-041) and under the auspices of the US Department of Energy by Lawrence Livermore National Laboratory (LLNL) under Contract No. DE-AC52- 07NA27344. Part of the funding was provided through the LLNL Livermore Graduate Scholar Program. Portions of this work were performed at HPCAT (Sector 16), Advanced Photon Source (APS), Argonne National Laboratory. HPCAT operation is supported by DOE-NNSA under Award No. DE-NA0001974, with partial instrumentation funding by NSF. The Advanced Photon Source is a U.S. Department of Energy (DOE) Office of Science User Facility operated for the DOE Office of Science by Argonne National Laboratory under Contract No. DE-AC02-06CH11357. P.C., Y.X., R.F. and C.K-B. acknowledge the support of DOE-BES/DMSE under Award DE-FG02-99ER45775. This material is based upon work supported by the NSF under Grant Number NSF DMR-1609855.